\documentclass[pra,aps,showpacs,twocolumn,longbibliography]{revtex4}
\usepackage{amssymb}
\usepackage{amsmath}
\usepackage{mathtools}
\usepackage[dvipdfm]{hyperref}
\usepackage{hyperref}
\usepackage{stmaryrd}
\usepackage{float}

\allowdisplaybreaks
\begin{document}

\title{Classification of incompatibility for two orthonormal bases}
\author{Jianwei Xu}
\email{xxujianwei@nwafu.edu.cn}
\affiliation{College of Science, Northwest A$\&$F University, Yangling, Shaanxi 712100,
China}

\begin{abstract}
For two orthonormal bases of a $d$-dimensional complex Hilbert space, the notion of
complete incompatibility was introduced recently by De Bi\`{e}vre [Phys. Rev. Lett. 127, 190404 (2021)]. In this work, we introduce
the notion of $s$-order incompatibility with positive integer $s$ satisfying
$2\leq s\leq d+1.$ In particular, $(d+1)$-order incompatibility just
coincides with the complete incompatibility. We establish some relations
between $s$-order incompatibility, minimal support uncertainty and
rank deficiency of the transition matrix.  As an example, we determine the incompatibility order of the discrete Fourier transform with any finite dimension.
\end{abstract}

\maketitle


\section{Introduction}

Quantum physics manifests many properties different from classical physics,
these properties are called quantum nonclassicality. There are diverse
aspects and notions of quantum nonclassicality, such as noncommutativity of
two operators, entanglement, coherence, uncertainty principles, nonreality,
contextuality, and nonlocality. These nonclassical properties remarkably
deepened the understanding of quantum physics and provided fruitful
applications in quantum technology.

Suppose $A=\{|a_{j}\rangle \}_{j=1}^{d}$, $B=\{|b_{k}\rangle \}_{k=1}^{d}$
are two orthonormal bases of a $d$-dimensional complex Hilbert space $H.$ To
avoid the freedom $|a_{j}\rangle \rightarrow e^{i\theta _{j}}|a_{j}\rangle $
with $\theta _{j}\in R$ real number and $i=\sqrt{-1}$, we denote $\overline{A%
}=\{|a_{j}\rangle \langle a_{j}|\}_{j=1}^{d}$ and $\overline{B}%
=\{|b_{k}\rangle \langle b_{k}|\}_{k=1}^{d},$ that is, $\overline{A}$ and $%
\overline{B}$ are all rank-1 projective measurements. We adopt the notion
``incompatibility'' as in Ref. \cite{PRL-2021-Bievre} that, when $\overline{A%
} $ and $\overline{B}$ commute we say $A$ and $B$ are compatible, otherwise
we say $A$ and $B$ are incompatible. $\overline{A}$ and $\overline{B}$
commute means that $|a_{j}\rangle \langle a_{j}|$ and $|b_{k}\rangle \langle
b_{k}|$ commute for any $j,k\in \llbracket{1,d}\rrbracket$ with $%
\llbracket{1,d}\rrbracket$ represents the set of consecutive integers $%
\{j\}_{j=1}^{d}.$ Thus $A$ and $B$ are compatible iff (if and only if) $%
\overline{A}=\overline{B}.$

The term \textquotedblleft incompatible" in the literature usually refers to
the meaning that two positive operator-valued measures (POVMs) are not
jointly measurable, such as in Refs. \cite%
{JMP-2010-Wolf,PRA-2015-Heinosaari,JMP-2017-Haapasalo,PRL-2019-Carmeli,NJP-2019-Designolle,PRA-2019-Zhang,PRA-2020-Martins,EPL-2020-Carmeli,PRL-2020-Zhou, PRA-2021-Brunner,PRR-2021-Chen}, and recent reviews see \cite{arXiv-2021-Otfried,JPA-2016-Heinosaari}. A POVM $D$ can be expressed by a set of positive semidefinite operators $%
D=\{D_{j}\}_{j=1}^{m}$ which sum to unity. Two POVMs $%
D=\{D_{j}\}_{j=1}^{m}$ and $E=\{E_{k}\}_{k=1}^{n}$ are called compatible iff
there exists a POVM $G=\{G_{jk}\}_{j=1}^{m},_{k=1}^{n}$ such that $\sum
{}_{j=1}^{m}G_{jk}=E_{k}$ for any $k,$ and $\sum {}_{k=1}^{n}G_{jk}=D_{j}$
for any $j.$ As a special case, when two measurements are two rank-$1$
projective measurements ($\overline{A},\overline{B}$) above, we can check
that ($\overline{A},\overline{B}$) are jointly measurable iff $\overline{A}=%
\overline{B}$. In this work, we only consider the incompatibility of two
rank-$1$ projective measurements ($\overline{A},\overline{B}$). Notice that in some works the term \textquotedblleft incompatible" may refer to different meanings than joint measurable. For example, in Ref. \cite{JPA-2022-Kaniewski} the notion of compatibility corresponds to commutativity
of the measurement operators.

In Ref. \cite{PRL-2021-Bievre}, De Bi\`{e}vre introduced the notion of
complete incompatibility. Two orthonormal bases $A=\{|a_{j}\rangle
\}_{j=1}^{d}$, $B=\{|b_{k}\rangle \}_{k=1}^{d}$ are completely incompatible,
if for any nonempty subsets $\varnothing \neq S_{A}\subseteq A$, $%
\varnothing \neq S_{B}\subseteq B$, $|S_{A}|+|S_{B}|\leq d$, it holds that
span$\{S_{A}\}\cap $span$\{S_{B}\}=\{0\}.$ Here $|S_{A}|$ stands for
the number of elements in $S_{A},$ span$\{S_{A}\}$ is the subspace spanned
by $S_{A}$ over the complex field $C.$ Although the definition of complete
incompatibility is purely algebraic, it possesses the physical
interpretation in terms of selective projective measurements \cite%
{PRA-2007-Johansen, Book-Nielsen-2010, Book-Claude-2015}. It is shown that
complete incompatibility closely links with the minimal support uncertainty
\cite{PRL-2021-Bievre}, and also, it is useful to characterize the
Kirkwood-Dirac nonclassicality \cite{PRL-2021-Bievre}.

In this work, we introduce the notion of $s$-order incompatibility with $%
s\in \llbracket{2,d+1}\rrbracket.$ Under this definition, complete
incompatibility is just $(d+1)$-order incompatibility. This paper is
organized as follows. In section II, we give the definition of $s$-order
incompatibility, and establish a link between it and the minimal support
uncertainty. In section III, we characterize $s$-order incompatibility via
the transition matrix of the two orthonormal bases. In section IV, we give examples to illustrate the calculation of incompatibility
order. Section V is a brief summary.

\section{$s$-order incompatibility and minimal support uncertainty}

We give the definition of $s$-order incompatibility, and establish a
relation between it and the minimal support uncertainty.

\emph{Definition 1. }$s$-order incompatibility. Suppose the integer $s$
satisfies $s\in \llbracket{2,d+1}\rrbracket$, $A=\{|a_{j}\rangle
\}_{j=1}^{d} $ and $B=\{|b_{k}\rangle \}_{k=1}^{d}$ are two orthonormal
bases of $d$-dimensional complex Hilbert space $H.$ We say $A$ and $B$ are $%
s $-order incompatible if the following (1.1) and (1.2) hold.

(1.1). For any $\varnothing \neq S_{A}\subseteq A$ and $\varnothing \neq
S_{B}\subseteq B$, if $|S_{A}|+|S_{B}|<s,$ then span$\{S_{A}\}\cap $ span$%
\{S_{B}\}=\{0\}.$

(1.2). There exist $\varnothing \neq S_{A}\subseteq A$ and $\varnothing \neq
S_{B}\subseteq B$, such that $|S_{A}|+|S_{B}|=s$ and span$\{S_{A}\}\cap $
span$\{S_{B}\}\neq \{0\}.$

We use $\chi _{AB}$ to denote the incompatibility order of $A$ and $B.$ When
$\chi _{AB}=d+1$, the $(d+1)$-order incompatibility just coincides with the
complete incompatibility introduced in Ref. \cite{PRL-2021-Bievre}.

We establish a link between $s$-order incompatibility and the minimal
support uncertainty. For a pure state $|\psi \rangle ,$ we express it in the
orthonormal bases $A=\{|a_{j}\rangle \}_{j=1}^{d}$ and $B=\{|b_{k}\rangle
\}_{k=1}^{d}$ as $|\psi \rangle =\sum_{j=1}^{d}|a_{j}\rangle \langle
a_{j}|\psi \rangle $ and $|\psi \rangle =\sum_{k=1}^{d}|b_{k}\rangle \langle
b_{k}|\psi \rangle .$ We use $n_{A}(|\psi \rangle )$ to denote the number of
nonzero elements in $\{\langle a_{j}|\psi \rangle \}_{j=1}^{d}$, use $%
n_{B}(|\psi \rangle )$ to denote the number of nonzero elements in $%
\{\langle b_{k}|\psi \rangle \}_{k=1}^{d},$ and let
\begin{eqnarray}
n_{AB}(|\psi \rangle ) &:&=n_{A}(|\psi \rangle )+n_{B}(|\psi \rangle ), \label{eq1-1}
\label{eq1} \\
n_{AB}^{\min } &:&=\min_{|\psi \rangle \neq 0}n_{AB}(|\psi \rangle ).  \label{eq1-2}
\label{eq2}
\end{eqnarray}
$n_{AB}(|\psi \rangle )$ is called the support uncertainty of $|\psi \rangle
$ with respect to $A$ and $B,$ and $n_{AB}^{\min }$ is called the minimal
support uncertainty with respect to $A$ and $B.$ The
support uncertainty $n_{AB}(|\psi \rangle )$ has many applications in
different situations \cite{JAM-1989-Stark, AMUC-2004-Przebinda,
MRL-2005-Tao, LAA-2011-Jaming, Wigderson-2021-BAMS}. Obviously, $%
n_{AB}^{\min }\in \llbracket{2,d+1}\rrbracket.$ It is shown that $\chi
_{AB}=d+1$ iff $n_{AB}^{\min }=d+1$ \cite{PRL-2021-Bievre}. We now prove a
more general result in Theorem 2.

\emph{Theorem 2.} Suppose $A=\{|a_{j}\rangle \}_{j=1}^{d}$ and $%
B=\{|b_{k}\rangle \}_{k=1}^{d}$ are two orthonormal bases of $d$-dimensional
complex Hilbert space $H$. The incompatibility order $\chi _{AB}$ and
minimal support uncertainty $n_{AB}^{\min }$ are defined in Definition 1 and
Eq. (\ref{eq1-2}), then it holds that
\begin{eqnarray}
\chi _{AB}=n_{AB}^{\min }.  \label{eq1-3}
\end{eqnarray}
\emph{Proof.} By the definition of $n_{AB}^{\min }$, if $n_{AB}^{\min }=s,$
then there exists pure state $|\psi \rangle $ such that $n_{AB}(|\psi
\rangle )=n_{A}(|\psi \rangle )+n_{B}(|\psi \rangle )=s$ and there does not
exist pure state $|\psi ^{\prime }\rangle $ such that $n_{AB}(|\psi ^{\prime
}\rangle )=n_{A}(|\psi ^{\prime} \rangle )+n_{B}(|\psi ^{\prime} \rangle )<s.$ For such $|\psi
\rangle ,$ there exist $\varnothing \neq S_{A}\subseteq A$ and $\varnothing
\neq S_{B}\subseteq B$, such that $|S_{A}|=n_{A}(|\psi \rangle )$, $%
|S_{B}|=n_{B}(|\psi \rangle )$, and $|\psi \rangle \in $span$\{S_{A}\}\cap $
span$\{S_{B}\}.$ The nonexistence of such $|\psi ^{\prime }\rangle $ implies
that there does not exist $\varnothing \neq S_{A}\subseteq A$ and $%
\varnothing \neq S_{B}\subseteq B$, such that $|S_{A}|=n_{A}(|\psi ^{\prime
}\rangle )$, $|S_{B}|=n_{B}(|\psi ^{\prime }\rangle )$, and $|\psi ^{\prime
}\rangle \in $span$\{S_{A}\}\cap $ span$\{S_{B}\}.$ These two conditions
just coincide with (1.1) and (1.2) in Definition 1. Then the claim follows. $%
\hfill\blacksquare$

Again, when $s=d+1,$ Theorem 1 returns to the corresponding result in Ref.
\cite{PRL-2021-Bievre}.

\section{$s$-order incompatibility and the transition matrix}

In this section, we introduce the index of rank deficiency $\tau _{AB}$. We also establish a
link between $\chi _{AB}$ ($n_{AB}^{\min }$) and $\tau _{AB}$, then $\chi
_{AB}$ can be determined via $\tau _{AB}.$

For two orthonormal bases $A=\{|a_{j}\rangle \}_{j=1}^{d}$ and $%
B=\{|b_{k}\rangle \}_{k=1}^{d},$ the transition matrix $%
U^{AB}=(U_{jk}^{AB})_{j,k=1}^{d}$ is defined as $U_{jk}^{AB}=\langle
a_{j}|b_{k}\rangle .$ Conversely, for a given unitary matrix $U$, we can always find two orthonormal bases $A=\{|a_{j}\rangle \}_{j=1}^{d}$ and $%
B=\{|b_{k}\rangle \}_{k=1}^{d}$ such that $U_{jk}=\langle
a_{j}|b_{k}\rangle .$ For example, when express $U=(U_{jk})_{j,k=1}^{d}$ in the standard computational basis $\{|j\rangle\}_{j=1}^{d}$, let $A$ be this standard computational basis and $B$ be the column vectors of $U=(U_{jk})_{j,k=1}^{d}$. Note that $U_{jk}^{AB}=\langle
a_{j}|b_{k}\rangle=\langle
a_{j}|V^{\dagger}V|b_{k}\rangle$  for any $d\times d$ unitary matrix $V$ with $V^{\dagger}$ the Hermitian conjugate of $V$, then the transition matrix $U$ with respect to $(A,B)$ is invariant under the unitary operation $V:(A,B)\rightarrow(VA,VB)$. Here $VA=\{V|a_{j}\rangle \}_{j=1}^{d}$.

 We want to characterize $s$-order incompatibility via
the transition matrix $U^{AB}.$ To do this, we introduce the definition of $%
t $-order rank deficiency of $U^{AB}.$

\emph{Definition 3.} $t$-order rank deficiency of $U^{AB}.$ For the
transition matrix $U^{AB}$ and the integer $t\in \llbracket{0,d-1}\rrbracket,
$ we define the $t$-order rank deficiency of $U^{AB}$ as follows.
\begin{eqnarray}
&&R_{t,r}(U^{AB})  \notag \\
&=&\max_{\substack{ 1\leq m\leq d-t; \\ 1\leq j_{1}<j_{2}<...<j_{m}\leq d;
\\ 1\leq k_{1}<k_{2}<...<k_{m+t}\leq d}}\{m-\text{rank}\binom{%
j_{1},j_{2},...,j_{m};}{k_{1},k_{2},...,k_{m+t}.}\},  \notag \\  \label{eq2-1} \\
&&R_{t,c}(U^{AB})  \notag \\
&=&\max_{_{\substack{ 1\leq m\leq d-t; \\ 1\leq j_{1}<j_{2}<...<j_{m}\leq d;
\\ 1\leq k_{1}<k_{2}<...<k_{m+t}\leq d}}}\{m-\text{rank}\binom{%
k_{1},k_{2},...,k_{m+t};}{j_{1},j_{2},...,j_{m}.}\},  \notag \\  \label{eq2-2} \\
&&R_{t}(U^{AB})=\max \{R_{t,r}(U^{AB}),R_{t,c}(U^{AB})\}.  \label{eq2-3}
\end{eqnarray}%
Where $\binom{j_{1},j_{2},...,j_{m};}{k_{1},k_{2},...,k_{m+t}.}$ denotes the
submatrix obtained by the $(j_{1},j_{2},...,j_{m})$ rows and $%
(k_{1},k_{2},...,k_{m+t})$ columns of $U^{AB}$, for example $\binom{1,3;}{%
2,3,4.}=\left(
\begin{array}{ccc}
\langle a_{1}|b_{2}\rangle  & \langle a_{1}|b_{3}\rangle  & \langle
a_{1}|b_{4}\rangle  \\
\langle a_{3}|b_{2}\rangle  & \langle a_{3}|b_{3}\rangle  & \langle
a_{3}|b_{4}\rangle
\end{array}%
\right) .$

Clearly, the definitions of $R_{t,r}(U^{AB})$, $R_{t,c}(U^{AB}),$
and $R_{t}(U^{AB})$ above can be similarly defined for general matrices, not
only the unitary matrices. Note that a similar definition of rank-deficient
submatrices was proposed in Ref. \cite{Delvaux-2008-LAA}.

\emph{Proposition 4.} Suppose $t \in \llbracket{0,d-1}\rrbracket,$ then the following (4.1)-(4.4) hold.

(4.1). $R_{t}(U^{AB})\geq 0.$

(4.2). $0\leq R_{t}(U^{AB})-R_{t+1}(U^{AB})\leq 1.$

(4.3). $R_{d-1}(U^{AB})=0.$

(4.4). If $R_{0}(U^{AB})=0$ then $R_{t}(U^{AB})=0$ for any $t\in %
\llbracket{0,d-1}\rrbracket.$

\emph{Proof.} Recall that the matrix rank is defined as the rank of row
vectors and which also equals the rank of column vectors, then $%
R_{t}(U^{AB})\geq 0$ evidently holds since $m\geq $rank$\binom{%
j_{1},j_{2},...,j_{m};}{k_{1},k_{2},...,k_{m+t}.}$ and $m\geq $rank$\binom{k_{1},k_{2},...,k_{m+t};}{
j_{1},j_{2},...,j_{m}.}.$

For $t+1,$ according to Definition 3, there exist $1\leq m\leq d-(t+1)$ and $%
\binom{j_{1},j_{2},...,j_{m};}{k_{1},k_{2},...,k_{m+t+1}.}$ such that $%
R_{t+1}(U^{AB})=m-$rank$\binom{j_{1},j_{2},...,j_{m};}{%
k_{1},k_{2},...,k_{m+t+1}.}$, or there exist $1\leq n\leq (t+1)$ and $\binom{%
k_{1},k_{2},...,k_{n+t+1};}{j_{1},j_{2},...,j_{n}.}$ such that $%
R_{t+1}(U^{AB})=n-$rank$\binom{k_{1},k_{2},...,k_{n+t+1};}{%
j_{1},j_{2},...,j_{n}.}.$ We consider the former case, the latter can be
discussed similarly. For the former case, we see that
\begin{eqnarray*}
&&R_{t+1}(U^{AB})=m-\text{rank}\binom{j_{1},j_{2},...,j_{m};}{%
k_{1},k_{2},...,k_{m+t+1}.} \\
&\leq &(m+1)-\text{rank}\binom{l_{1},l_{2},...,l_{m},l_{m+1};}{%
k_{1},k_{2},...,k_{m+t+1}.} \\
&\leq &R_{t}(U^{AB}),
\end{eqnarray*}%
where $0<l_{1}<l_{2}<...<l_{m}<l_{m+1}\leq d$ and $\{j_{1},j_{2},...,j_{m}\}%
\subseteq \{l_{1},l_{2},...,l_{m},l_{m+1}\}.$ The first inequality says the
fact that adding one row can at most increase 1 for the rank. The second
inequality is from the definition of $R_{t}(U^{AB})$. Then $R_{t}(U^{AB})\geq R_{t+1}(U^{AB})$.

When $t=d-1,$ from Definition 3, $m$ can only take $m=1.$ Since $U^{AB}$ is
unitary, then every row vector and every column vector of $U^{AB}$ are all
nonzero. Hence, $R_{d-1}(U^{AB})=0.$ This proves (4.3).

(4.4) is a direct result of $R_{t}(U^{AB})\geq R_{t+1}(U^{AB})$ and (4.3).

Lastly, we prove $R_{t}(U^{AB})-R_{t+1}(U^{AB})\leq 1$. If $R_{t}(U^{AB})\leq 1$ then the claim is obviously true. Suppose $R_{t}(U^{AB})\geq 2$ and the submatrix $\binom{j_{1},j_{2},...,j_{m};}{k_{1},k_{2},...,k_{m+t}.}$ reaches  $R_{t}(U^{AB})=m-$rank$\binom{j_{1},j_{2},...,j_{m};}{k_{1},k_{2},...,k_{m+t}.}$, we see that $m \geq 2$. Removing any row, the remaining submatrix, for example, is $\binom{j_{1},j_{2},...,j_{m-1};}{k_{1},k_{2},...,k_{m+t}.}.$  We have that
\begin{eqnarray*}
&&R_{t+1}(U^{AB}) \\
&\geq& (m-1)-\text{rank}\binom{j_{1},j_{2},...,j_{m-1};}{%
k_{1},k_{2},...,k_{m+t}.} \\
&\geq &m-\text{rank}\binom{j_{1},j_{2},...,j_{m};}{%
k_{1},k_{2},...,k_{m+t}.}-1 \\
&=&R_{t}(U^{AB})-1,
\end{eqnarray*}%
 then (4.2) is true and we finished this proof. $\hfill \blacksquare $

With Proposition 4, we propose the definition of the index of rank
deficiency of the transition matrix $U^{AB}.$

\emph{Definition 5.} We define the index of rank deficiency of the
transition matrix $U^{AB}$ as
\begin{equation}
\tau _{AB}:=\min_{t\in \llbracket{0,d-1}\rrbracket}\{t\Big|%
R_{t}(U^{AB})=0\}-1.  \label{eq2-4}
\end{equation}%

Clearly, $\tau _{AB}\in \llbracket{-1,d-2}\rrbracket.$ When $R_{0}(U^{AB})=0,
$ we have that $\tau _{AB}=-1.$ For $\tau _{AB}=-1,$ every $m\times m$
submatrix $\binom{j_{1},j_{2},...,j_{m};}{k_{1},k_{2},...,k_{m}.}$ is of
rank $m,$ particularly, every element $U_{jk}^{AB}=\langle
a_{j}|b_{k}\rangle \neq 0.$ When $\tau _{AB}\in \llbracket{0,d-2}\rrbracket,$
$\tau _{AB}$ is the maximal $t$ for which $R_{t}(U^{AB})>0$, for such case
it must hold $R_{\tau _{AB}}(U^{AB})=1.$  Hence we have Corollary 1 below.

\emph{Corollary 1.} Suppose $\tau _{AB}\in \llbracket{0,d-2}\rrbracket,$
then $R_{\tau _{AB}}(U^{AB})=1,$ and
\begin{equation}
\tau _{AB}=\max_{t\in \llbracket{0,d-1}\rrbracket}\{t\Big|R_{t}(U^{AB})=1\}.   \label{eq2-5}
\end{equation}

If $\binom{j_{1},j_{2},...,j_{m};}{k_{1},k_{2},...,k_{m+\tau _{AB}}.}$ reaches $R_{\tau _{AB}}(U^{AB})=1,$ we assert that there must exist $%
\{z_{j}\}_{j=1}^{m}$ being complex numbers and all nonzero such that
\begin{equation}
(z_{1},z_{2},...,z_{m})\binom{j_{1},j_{2},...,j_{m};}{%
k_{1},k_{2},...,k_{m+\tau _{AB}}.}=0.  \label{eq2-6}
\end{equation}
Otherwise, if $\{z_{j}\}_{j=1}^{m}$ are not all nonzero, for example, $\{z_{j}\neq 0\}_{j=1}^{m-1}$ and $%
z_{m}=0,$ then Eq. (\ref{eq2-6}) implies $\binom{j_{1},j_{2},...,j_{m-1};}{%
k_{1},k_{2},...,k_{m+\tau _{AB}}.}$ is rank deficient in rows and $%
R_{\tau _{AB}+1}(U^{AB})\geq 1$, this contradicts Eq. (\ref{eq2-5}).

Similarly, if $\binom{k_{1},k_{2},...,k_{n+\tau _{AB}};}{j_{1},j_{2},...,j_{n}.}$
reaches $R_{\tau _{AB}}(U^{AB})=1,$ then there exist $%
\{z_{j}\}_{j=1}^{n}$ being complex numbers and all nonzero such that
\begin{equation}
\binom{k_{1},k_{2},...,k_{n+\tau _{AB}};}{j_{1},j_{2},...,j_{n}.}%
(z_{1},z_{2},...,z_{n})^{t}=0,  \label{eq2-7}
\end{equation}
where $(\ \ )^{t}$ denotes the transpose.

In Ref. \cite{PRL-2021-Bievre}, it is shown that when $A$ and $B$ are
completely incompatible, i.e., $\chi _{AB}=d+1,$ then it holds that $\tau
_{AB}=-1.$ Theorem 6 below shows a more general result, which is the central result of this work.

\emph{Theorem 6.} Suppose $A=\{|a_{j}\rangle \}_{j=1}^{d}$ and $%
B=\{|b_{j}\rangle \}_{j=1}^{d}$ are two orthonormal bases of $d$-dimensional
complex Hilbert space $H$. Then the incompatibility order $\chi _{AB}$ and
the index of rank deficiency $\tau _{AB}$ have the relation
\begin{equation}
\chi _{AB}+\tau _{AB}=d.  \label{eq2-8}
\end{equation}%
\emph{Proof.} The case of $\chi _{AB}=d+1$ has been proved in Ref. \cite{PRL-2021-Bievre}, then we only consider the case of $2\leq \chi _{AB}\leq d.$  Suppose the incompatibility order is $\chi _{AB},$ then (1.2) in
Definition 1 holds, that is, there exist $\varnothing \neq S_{A}\subseteq A$
and $\varnothing \neq S_{B}\subseteq B$ such that $|S_{A}|+|S_{B}|=\chi _{AB}
$ and span$\{S_{A}\}\cap $ span$\{S_{B}\}\neq \{0\}.$ Then there exists a
pure state $|\psi \rangle \in $span$\{S_{A}\}\cap $ span$\{S_{B}\}.$ Without
loss of generality, we assume $S_{A}=\{|a_{j}\rangle \}_{j=1}^{|S_{A}|}$, $%
S_{B}=\{|b_{k}\rangle \}_{k=1}^{|S_{B}|}.$ We explicitly write $U^{AB}$ as
\begin{widetext}
\begin{eqnarray}
\left(
\begin{array}{c}
\begin{array}{ccc}
\langle a_{1}|b_{1}\rangle & ... & \langle a_{1}|b_{|S_{B}|}\rangle \\
... & ... & ... \\
\langle a_{|S_{A}|}|b_{1}\rangle & ... & \langle
a_{|S_{A}|}|b_{|S_{B}|}\rangle%
\end{array}%
\begin{array}{ccc}
\langle a_{1}|b_{|S_{B}|+1}\rangle & ... & \langle a_{1}|b_{d}\rangle \\
... & ... & ... \\
\langle a_{|S_{A}|}|b_{|S_{B}|+1}\rangle & ... & \langle
a_{|S_{A}|}|b_{d}\rangle%
\end{array}
\\
\begin{array}{ccc}
\langle a_{|S_{A}|+1}|b_{1}\rangle & ... & \langle
a_{|S_{A}|+1}|b_{|S_{B}|}\rangle \\
... & ... & ... \\
\langle a_{d}|b_{1}\rangle & ... & \langle a_{d}|b_{|S_{B}|}\rangle%
\end{array}%
\begin{array}{ccc}
\langle a_{|S_{A}|+1}|b_{|S_{B}|+1}\rangle & ... & \langle
a_{|S_{A}|+1}|b_{d}\rangle \\
... & ... & ... \\
\langle a_{d}|b_{|S_{B}|+1}\rangle & ... & \langle a_{d}|b_{d}\rangle%
\end{array}%
\end{array}%
\right).  \label{eq2-9}
\end{eqnarray}
\end{widetext}Expanding $|\psi \rangle $ in $S_{A}$ and $S_{B}$ we get that
\begin{equation}
|\psi \rangle =\sum_{j=1}^{|S_{A}|}x_{j}|a_{j}\rangle
=\sum_{k=1}^{|S_{B}|}y_{k}|b_{k}\rangle ,  \notag
\end{equation}%
where $\{x_{j}\}_{j=1}^{S_{A}}$ are  all nonzero complex numbers, $%
\{y_{k}\}_{k=1}^{S_{B}}$ are  all nonzero complex numbers. $x_{j}=0$ or $y_{k}=0$ will contradicts $|S_{A}|+|S_{B}|=\chi _{AB}=n_{AB}^{\text{min}}.$ Consequently,
\begin{eqnarray}
\langle \psi |b_{k}\rangle  &=&0\text{ for all }|S_{B}|+1\leq k\leq d,
\notag  \label{eq12} \\
\langle a_{j}|\psi \rangle  &=&0\text{ for all }|S_{A}|+1\leq j\leq d.
\notag  \label{eq13}
\end{eqnarray}%
These imply that
\begin{eqnarray}
\sum_{j=1}^{|S_{A}|}x_{j}^{\ast }\langle a_{j}|b_{k}\rangle  &=&0\text{ for
all }|S_{B}|+1\leq k\leq d,  \notag   \\
\sum_{k=1}^{|S_{B}|}y_{k}\langle a_{j}|b_{k}\rangle  &=&0\text{ for all }%
|S_{A}|+1\leq j\leq d,  \notag
\end{eqnarray}%
where $x_{j}^{\ast }$ is the complex conjugate of $x_{j}.$ These say that
the $|S_{A}|\times (d-|S_{B}|)$ submatrix $\binom{1,2,...,|S_{A}|;}{%
|S_{B}|+1,|S_{B}|+2,...,d.}$ has linearly dependent row vectors, and the $%
(d-|S_{A}|)\times |S_{B}|$ submatrix $\binom{|S_{A}|+1,|S_{A}|+2,...,d;}{%
1,2,...,|S_{B}|.}$ has linearly dependent column vectors. Since $2\leq \chi
_{AB}\leq d,$ then $|S_{A}|+|S_{B}|\leq d$, $|S_{A}|\leq d-|S_{B}|$, and $%
|S_{B}|\leq d-|S_{A}|.$ These further imply that $%
R_{d-|S_{A}|-|S_{B}|}(U^{AB})>0$ and $\tau _{AB}\geq d-\chi _{AB}.$

Conversely, suppose the index of rank deficiency is $\tau _{AB}$, then there
exist $1\leq m\leq d-\tau _{AB}$ and $\binom{j_{1},j_{2},...,j_{m};}{%
k_{1},k_{2},...,k_{m+\tau _{AB}}.}$ such that $m-$rank$\binom{%
j_{1},j_{2},...,j_{m};}{k_{1},k_{2},...,k_{m+\tau _{AB}}.}=1$, or there
exist $1\leq n\leq d-\tau _{AB}$ and $\binom{k_{1},k_{2},...,k_{n+\tau
_{AB}};}{j_{1},j_{2},...,j_{n}.}$ such that $n-$rank$\binom{%
k_{1},k_{2},...,k_{n+t_{2}};}{j_{1},j_{2},...,j_{n}.}=1.$ We consider the
former case, the latter can be discussed similarly. For the former case,
without loss of generality, we assume $\binom{j_{1},j_{2},...,j_{m};}{%
k_{1},k_{2},...,k_{m+\tau _{AB}}.}=\binom{1,2,...,m;}{1,2,...,m+\tau _{AB}.}.
$ Since $m-$rank$\binom{1,2,...,m;}{1,2,...,m+\tau _{AB}.}=1,$  then there must exist $\{z_{j}\}_{j=1}^{m}$ being complex numbers
and all nonzero such that
\begin{equation}
\sum_{j=1}^{m}z_{j}\langle a_{j}|b_{k}\rangle =0\text{ for all }1\leq k\leq
m+\tau _{AB}.  \notag
\end{equation}%
Let $|\varphi \rangle =\sum_{j=1}^{m}z_{j}^{\ast }|a_{j}\rangle $, then $%
|\varphi \rangle \neq 0$ and
\begin{equation}
\langle \varphi |b_{k}\rangle =0\text{ for all }1\leq k\leq m+\tau _{AB}.
\notag
\end{equation}%
It follows that $n_{A}(|\varphi\rangle )=m$, $n_{B}(|\varphi\rangle )\leq d-m-\tau _{AB},$
and
\begin{equation}
\chi _{AB}=n_{AB}^{\min }\leq n_{A}(|\varphi\rangle )+n_{B}(|\varphi\rangle )\leq d-\tau
_{AB}.  \notag
\end{equation}%
Theorem 6 then follows. $\hfill\blacksquare$

Theorem 6 and Theorem 2 provide a way to determine $\chi _{AB}$ and $%
n_{AB}^{\min }$ via $\tau _{AB}.$ From the proof of Theorem 6, we see that
there exist $\binom{j_{1},j_{2},...,j_{m};}{k_{1},k_{2},...,k_{m+1}.}$ and $%
\binom{k_{1},k_{2},...,k_{n+1};}{j_{1},j_{2},...,j_{n}.}$ such that $m-$rank$%
\binom{j_{1},j_{2},...,j_{m};}{k_{1},k_{2},...,k_{m+1}.}=n-$rank$\binom{%
k_{1},k_{2},...,k_{n+1};}{j_{1},j_{2},...,j_{n}.}=1.$ We conclude this fact
as Corollary 2 below.

\emph{Corollary 2.} Suppose $\tau _{AB}\in \llbracket{0,d-2}\rrbracket,$ then
\begin{equation}
R_{\tau _{AB}}(U^{AB})=R_{\tau _{AB},r}(U^{AB})=R_{\tau _{AB},c}(U^{AB})=1. \label{eq2-10}
\end{equation}

\section{Examples}

We give some examples to illustrate the computation of $R_{t}(U^{AB})$ and
incompatibility order.

\emph{Example 1.} For $d=6,$ we consider $R_{t}(U^{AB})$ of the unity matrix
\begin{equation}
U^{AB}=I_{6}=\left(
\begin{array}{cccccc}
1 & 0 & 0 & 0 & 0 & 0 \\
0 & 1 & 0 & 0 & 0 & 0 \\
0 & 0 & 1 & 0 & 0 & 0 \\
0 & 0 & 0 & 1 & 0 & 0 \\
0 & 0 & 0 & 0 & 1 & 0 \\
0 & 0 & 0 & 0 & 0 & 1%
\end{array}%
\right) .
\end{equation}

$I_{6}$ is symmetric, then $R_{t}(I_{6})=R_{t,r}(I_{6})=R_{t,c}(I_{6}).$ We
only need to consider $R_{t,r}(I_{6})$. By (4.3) one sees that $%
R_{5}(I_{6})=0.$

For $R_{4}(I_{6}),$ (4.2) implies $R_{4}(I_{6})=0$ or $1.$ Since the
submatrix rank$\binom{1;}{2,3,4,5,6.}=0,$ then we get $R_{4}(I_{6})=1.$

For $R_{3}(I_{6}),$ (4.2) implies that $R_{3}(I_{6})=1$ or $2$. rank$\binom{%
j_{1};}{k_{1},k_{2},k_{3},k_{4}.}=0$ or $1,$ then $1-$rank$\binom{j_{1};}{%
k_{1},k_{2},k_{3},k_{4}.}=0$ or $1.$ rank$\binom{j_{1},j_{2},j_{3};}{%
1,2,3,4,5,6.}=3$, then $3-$rank$\binom{j_{1},j_{2},j_{3};}{1,2,3,4,5,6.}=0.$
rank$\binom{j_{1},j_{2};}{k_{1},k_{2},k_{3},k_{4},k_{5}.}=1$ or $2$, then $2-
$rank$\binom{j_{1},j_{2},j_{3};}{1,2,3,4,5,6.}=0$ or $1.$ In conclusion, $%
R_{3}(I_{6})=1.$

For $R_{2}(I_{6}),$ (4.2) implies that $R_{2}(I_{6})=1$ or $2$. Since rank$%
\binom{1,2;}{3,4,5,6.}=0$ and $2-$rank$\binom{1,2;}{3,4,5,6.}=2,$ then $%
R_{2}(I_{6})=2.$

For $R_{1}(I_{6}),$ (4.2) implies that $R_{1}(I_{6})=2$ or $3$. rank$\binom{%
j_{1},j_{2},j_{3};}{k_{1},k_{2},k_{3},k_{4}.}\in \{1,2,3\}$, then $3-$rank$%
\binom{j_{1},j_{2},j_{3};}{1,2,3,4,5,6.}\in \{0,1,2\}.$ rank$\binom{%
j_{1},j_{2},j_{3},j_{4};}{k_{1},k_{2},k_{3},k_{4},k_{5}.}=3$ or $4$, then $4-
$rank$\binom{j_{1},j_{2},j_{3};}{1,2,3,4,5,6.}=0$ or $1.$ rank$\binom{%
j_{1},j_{2},j_{3},j_{4},j_{5};}{1,2,3,4,5,6.}=5$, then $5-$rank$\binom{%
j_{1},j_{2},j_{3},j_{4},j_{5};}{1,2,3,4,5,6.}=0.$ In conclusion, $%
R_{1}(I_{6})=2.$

For $R_{0}(I_{6}),$ (4.2) implies that $R_{1}(I_{6})=2$ or $3$. Since rank$%
\binom{1,2,3;}{4,5,6.}=0$, $3-$rank$\binom{1,2,3;}{4,5,6.}=3,$ then $%
R_{0}(I_{6})=3.$

We depict $\{R_{t}(I_{6})\}_{t\in \llbracket{0,5}\rrbracket}$ in Figure 1.
As a result, $\tau _{AB}=4$, $\chi _{AB}=2.$

\begin{figure}
\includegraphics[width=8cm]{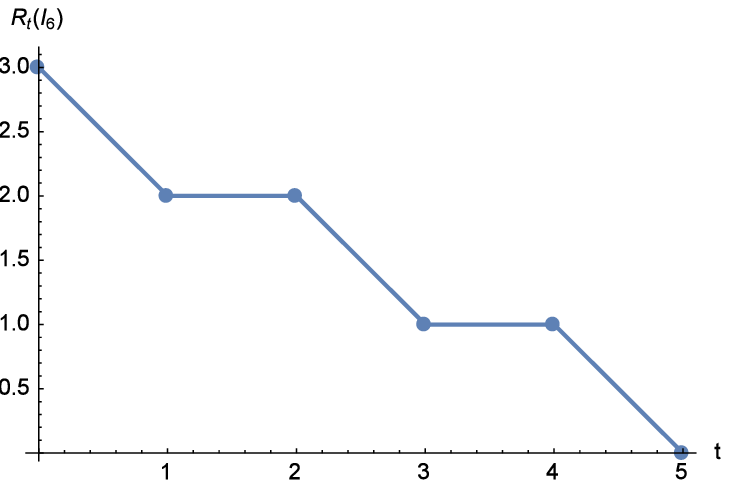}
\caption{Plot of $\{R_{t}(I_{6})\}_{t\in \llbracket{0,5}\rrbracket}$ in Example 1.}
\end{figure}

\emph{Example 2.} For qubit system, $d=2,$
\begin{eqnarray}
U^{AB}=\left(
\begin{array}{cc}
e^{i\varphi_{1}} \sin \theta & -e^{-i\varphi_{2}} \cos \theta \\
e^{i\varphi_{2}} \cos \theta & e^{-i\varphi_{1}} \sin \theta
\end{array}%
\right),
\end{eqnarray}
with $\theta, \varphi_{1}, \varphi_{2}$ real numbers and $\theta \in \lbrack 0,\frac{\pi }{2}].$
By (4.3) one sees that $R_{1}(U^{AB})=0.$
For $R_{0}(U^{AB}),$ (4.2) implies $R_{0}(U^{AB})=0$ or $1.$
When $\theta =0$ or $\frac{\pi }{2}$, $\sin \theta =0$ or $\cos \theta =0,$
we have $R_{0}(U^{AB})=1$, $\tau _{AB}=0$, $\chi _{AB}=2$, $\overline{A}=\overline{B}.$ When $%
0\neq \theta \neq \frac{\pi }{2}$, we have $R_{0}(U^{AB})=0$, $\tau _{AB}=-1$, $\chi _{AB}=3=d+1
$, $A$ and $B$ are completely incompatible.

\emph{Example 3.} For $d=3,$ consider the unitary matrix \cite{PRD-1988-Bronzan}
\begin{equation}
U^{AB}(\theta _{1},\theta _{2})=\left(
\begin{array}{ccc}
\cos \theta _{1}\cos \theta _{2} & \sin \theta _{1} & \cos \theta _{1}\sin
\theta _{2} \\
-\sin \theta _{1}\cos \theta _{2} & \cos \theta _{1} & -\sin \theta _{1}\sin
\theta _{2} \\
-\sin \theta _{2} & 0 & \cos \theta _{2}%
\end{array}%
\right) ,
\end{equation}
$\theta _{1},\theta _{2}\in \lbrack 0,\frac{\pi }{2}].$

By (4.3) one sees that $R_{2}(U^{AB})=0.$

For $R_{0}(U^{AB}),$ one sees that rank$\binom{1,2,3;}{1,2,3.}=3$, then $3-$%
rank$\binom{1,2,3;}{1,2,3.}=0.$ The unitarity of $U^{AB}$ implies that rank$%
\binom{j_{1},j_{2};}{k_{1},k_{2}.}\neq 0$, then rank$\binom{j_{1},j_{2};}{%
k_{1},k_{2}.}=1$ or 2, $2-$rank$\binom{j_{1},j_{2};}{k_{1},k_{2}.}=0$ or $1.$
Since rank$\binom{3;}{2.}=0$, then $1-$rank$\binom{3;}{2.}=1.$ Consequently,
$R_{0}(U^{AB})=1.$

For $R_{1}(U^{AB}),$ the unitarity of $U^{AB}$ implies that rank$\binom{%
j_{1},j_{2};}{1,2,3.}=2$, rank$\binom{1,2,3;}{k_{1},k_{2}.}=2$, then $2-$rank$\binom{j_{1},j_{2};}{1,2,3.}=0,$ $2-$rank$\binom{1,2,3;}{k_{1},k_{2}.}=0.$ rank%
$\binom{j_{1};}{k_{1},k_{2}.}=0$ or $1$, rank$\binom{j_{1},j_{2};}{k_{1}.}=0$
or $1$, and there exists rank$\binom{j_{1};}{k_{1},k_{2}.}=0$ or rank$\binom{j_{1},j_{2};%
}{k_{1}.}=0$ iff
\begin{equation}
\theta _{1}=0\text{ or }\theta _{1}=\frac{\pi }{2}\text{ or }\theta _{2}=0%
\text{ or }\theta _{2}=\frac{\pi }{2}.   \label{eq3-10}
\end{equation}
When Eq. (\ref{eq3-10}) holds then $R_{1}(U^{AB})=1,$ otherwise $R_{1}(U^{AB})=0.$

It follows that when Eq. (\ref{eq3-10}) holds then $\tau _{AB}=1$, $\chi _{AB}=2,$
otherwise $\tau _{AB}=0$, $\chi _{AB}=3.$

\emph{Example 4.} Discrete Fourier transform (DFT) matrix $F$. $F=U^{AB}$ is defined as $%
F_{jk}=U_{jk}^{AB}=\langle a_{j}|b_{k}\rangle =\frac{1}{\sqrt{d}}e^{i\frac{%
2\pi }{d}jk},$ with $i=\sqrt{-1}$, $j\in \llbracket{0,d-1}\rrbracket$, $k\in %
\llbracket{0,d-1}\rrbracket$.

It is shown that $A$, $B$ are completely incompatible ($\chi _{AB}=d+1$) iff $d$ is a prime \cite{MRL-2005-Tao,PRL-2021-Bievre}. We now consider the
general case that $d$ is not necessarily a prime. We have Theorem 7 below.

\emph{Theorem 7.} For $d$-dimensional DFT, it holds that
\begin{eqnarray}
\chi _{AB} &=&d^{\prime }+d/d^{\prime },    \label{eq3-5}  \\
d^{\prime }:&=&\max \{d_{1}\Big|d_{1}|d,d_{1}\leq \sqrt{d}\},  \label{eq3-6}
\end{eqnarray}
where $d_{1}|d$ means $d_{1}$ is a divisor of $d$.
We will provide a proof for Theorem 7 in Appendix.

We give another equivalent expression for Eq. (\ref{eq3-5}). Suppose $f$ is a nonzero
complex valued function on the index set $\{j\}_{j=0}^{d-1},$ let $\widehat{f}$ denote
the DFT of $f,$ that is,
\begin{equation*}
\widehat{f}(k)=\frac{1}{\sqrt{d}}\sum_{j=0}^{d-1}e^{i\frac{2\pi }{d}jk}f(j).
\end{equation*}
The support of $f,$ denoted by supp$f$, is defined as
\begin{equation*}
\text{supp}f:=\{j \Big| j\in \llbracket{0,d-1}\rrbracket \},f(j)\neq 0\}.
\end{equation*}
Let the pure state $|\psi \rangle =\sum_{j=0}^{d-1}f(j)|a_{j}\rangle ,$ then
\begin{equation*}
|\text{supp}f|=n_{A}(|\psi \rangle ).
\end{equation*}
Rewrite $|\psi \rangle =\sum_{j=0}^{d-1}f(j)|a_{j}\rangle
=\sum_{j,k=0}^{d-1}f(j)|b_{k}\rangle \langle b_{k}|a_{j}\rangle
=\sum_{j,k=0}^{d-1}F_{kj}f(j)|b_{k}\rangle =\sum_{j,k=0}^{d-1}\widehat{f}%
(k)|b_{k}\rangle ,$ then
\begin{equation*}
|\text{supp}\widehat{f}|=n_{B}(|\psi \rangle ).
\end{equation*}
We then can recast Eq. (\ref{eq3-5}) as an uncertainty principle
\begin{equation}
|\text{supp}f|+|\text{supp}\widehat{f}|\geq d^{\prime }+d/d^{\prime },  \label{eq3-7}
\end{equation}
and the lower bound on the right-hand side is sharp.

 In 1989, Donoho and Stark \cite{JAM-1989-Stark} established an
uncertainty principle for $|$supp$f|$ and $|$supp$\widehat{f}|$ of
DFT as
\begin{equation}
|\text{supp}f||\text{supp}\widehat{f}|\geq d,  \label{eq3-8}
\end{equation}
and the lower bound on the right-hand side is sharp.

In 2005, Tao \cite{MRL-2005-Tao} proved a stronger uncertainty principle of DFT
for $d=p$ with $p$ a prime, as
\begin{equation}
|\text{supp}f|+|\text{supp}\widehat{f}|\geq p+1,  \label{eq3-9}
\end{equation}
and the lower bound on the right-hand side is sharp.

We see that our result in Eq. (\ref{eq3-7}) evidently includes Eq. (\ref{eq3-9}) as a special case.

\section{Summary}

For two orthonormal bases $A,B$ of a quantum system, we introduced the notion
of incompatibility order $\chi _{AB}$, which resulted in a classification
for incompatibility. We introduced the notion of the index of rank deficiency
of the transition matrix $U^{AB}$, denoted by $\tau _{AB}$. We established a
link between $\chi _{AB}$ and minimal support uncertainty $n_{AB}^{\min },$
and established a link between $\chi _{AB}$ and $\tau _{AB}$. As an
application of these relations, we derived the incompatibility order of DFT.

\section*{ACKNOWLEDGMENTS}

This work was supported by the Natural Science Basic Research Plan in
Shaanxi Province of China (Program No. 2022JM-012). I thank the anonymous referees for constructive comments. I also thank Wen-Zhi Jia and Chang-Yong Liu for helpful discussions. After completing this work, I became aware of the recent work \cite{Bievre-2022-arXiv} which provides an in-depth study of the completely incompatibility and its links to the support uncertainty and to the Kirkwood-Dirac nonclassicality of pure quantum states.

\section*{Appendix: Proof of Theorem 7}

\setcounter{equation}{0} \renewcommand\theequation{A\arabic{equation}}

When $d$ is a prime, Theorem 7 returns to Eq. (\ref{eq3-9}) in main text, which has
been proved in \cite{MRL-2005-Tao}. Then we only consider the case that $d$
is not prime. Note that $F=F^{t},$ thus $R_{t,r}(F)=R_{t,c}(F).$

Suppose
\begin{equation}
d=d_{1}d_{2},  \label{A1}
\end{equation}
with $d_{1}|d$, $d_{2}|d$ and $1<d_{1}\leq d_{2}<d.$ We rewrite the index
sets $\{j\}_{j=0}^{d-1}$ and $\{k\}_{k=0}^{d-1}$ as
\begin{eqnarray}
j &=&j_{0}+j^{\prime }d_{2},j_{0}\in \llbracket{0,d_{2}-1}\rrbracket%
,j^{\prime }\in \llbracket{0,d_{1}-1}\rrbracket,  \label{A2} \\
k &=&k_{0}+k^{\prime }d_{1},k_{0}\in \llbracket{0,d_{1}-1}\rrbracket%
,k^{\prime }\in \llbracket{0,d_{2}-1}\rrbracket,  \ \ \label{A3}
\end{eqnarray}%
then
\begin{equation}
F_{jk}=\frac{1}{\sqrt{d}}e^{i\frac{2\pi }{d}jk}=\frac{1}{\sqrt{d}}e^{i\frac{2\pi }{d}j_{0}k_{0}}e^{i\frac{2\pi }{d}
j_{0}k^{\prime }d_{1}}e^{i\frac{2\pi }{d}k_{0}j^{\prime }d_{2}},  \label{A4}
\end{equation}%
where we have used the fact $e^{i\frac{2\pi }{d}j^{\prime }k^{\prime
}d_{1}d_{2}}=1.$ As pointed out in Ref. \cite{Delvaux-2008-LAA}, Eq. (\ref%
{A4}) implies that
\begin{equation}
\text{rank}\left(
\begin{array}{c}
\{j_{0}+j^{\prime }d_{2}\}_{j^{\prime }\in \llbracket{0,d_{1}-1}\rrbracket};
\\
\{k_{0}+k^{\prime }d_{1}\}_{k^{\prime }\in \llbracket{0,d_{2}-1}\rrbracket}.%
\end{array}%
\right) =1  \label{A5}
\end{equation}%
since
\begin{equation}
\left(
\begin{array}{c}
\{j_{0}+j^{\prime }d_{2}\}_{j^{\prime }\in \llbracket{0,d_{1}-1}\rrbracket};
\\
\{k_{0}+k^{\prime }d_{1}\}_{k^{\prime }\in \llbracket{0,d_{2}-1}\rrbracket}.%
\end{array}%
\right) =\frac{1}{\sqrt{d}}e^{i\frac{2\pi }{d}%
j_{0}k_{0}}F_{k_{0}}^{t}F_{j_{0}},  \label{A6}
\end{equation}%
where we have denoted the row vector
\begin{equation}
F_{k_{0}}=(1,e^{i\frac{2\pi }{d}k_{0}d_{2}},e^{2i\frac{2\pi }{d}%
k_{0}d_{2}},...e^{(d_{1}-1)i\frac{2\pi }{d}k_{0}d_{2}}),  \label{A7}
\end{equation}%
and denoted the transpose of $F_{k_{0}}$ by $F_{k_{0}}^{t}$.

The submatrix $\left(
\begin{array}{c}
\{j_{0}+j^{\prime }d_{2}\}_{j^{\prime }\in \llbracket{0,d_{1}-1}\rrbracket};
\\
\{k_{0}+k^{\prime }d_{1}\}_{k_{0}\in \llbracket{1,d_{1}-1}\rrbracket%
,k^{\prime }\in \llbracket{0,d_{2}-1}\rrbracket}.%
\end{array}%
\right) $ can be viewed as the column union of the submatrices $\left\{
\left(
\begin{array}{c}
\{j_{0}+j^{\prime }d_{2}\}_{j^{\prime }\in \llbracket{0,d_{1}-1}\rrbracket};
\\
\{k_{0}+k^{\prime }d_{1}\}_{k^{\prime }\in \llbracket{0,d_{2}-1}\rrbracket}.%
\end{array}%
\right) \right\} _{k_{0}\in \llbracket{1,d_{1}-1}\rrbracket},$ thus the
column rank (and then the rank)
\begin{eqnarray}
\text{rank}\left(
\begin{array}{c}
\{j_{0}+j^{\prime }d_{2}\}_{j^{\prime }\in \llbracket{0,d_{1}-1}\rrbracket};
\\
\{k_{0}+k^{\prime }d_{1}\}_{k_{0}\in \llbracket{1,d_{1}-1}\rrbracket%
,k^{\prime }\in \llbracket{0,d_{2}-1}\rrbracket}.%
\end{array}%
\right) \leq d_{1}-1.  \notag \\    \label{A8}
\end{eqnarray}

Since $\left(
\begin{array}{c}
\{j_{0}+j^{\prime }d_{2}\}_{j^{\prime }\in \llbracket{0,d_{1}-1}\rrbracket};
\\
\{k_{0}+k^{\prime }d_{1}\}_{k_{0}\in \llbracket{1,d_{1}-1}\rrbracket%
,k^{\prime }\in \llbracket{0,d_{2}-1}\rrbracket}.%
\end{array}%
\right) $ has $d_{1}$ rows, thus $\left(
\begin{array}{c}
\{j_{0}+j^{\prime }d_{2}\}_{j^{\prime }\in \llbracket{0,d_{1}-1}\rrbracket};
\\
\{k_{0}+k^{\prime }d_{1}\}_{k_{0}\in \llbracket{1,d_{1}-1}\rrbracket%
,k^{\prime }\in \llbracket{0,d_{2}-1}\rrbracket}.%
\end{array}%
\right) $ is rank deficient for rows. By the definition of $\tau _{AB},$ it
follows that $\tau _{AB}\geq (d_{1}-1)d_{2}-d_{1}$, that is
\begin{equation}
\tau _{AB}\geq d-(d_{1}+d_{2}).  \label{A9}
\end{equation}%
Minimizing ${d_{1}+d_{2}}$ over all $d_{1}$ under Eq. (\ref{A1}) will yield
\begin{eqnarray}
\tau _{AB} &\geq &d-(d^{\prime }+d/d^{\prime }),  \label{A10} \\
d^{\prime } &:&=\max \{d_{1}\Big|1<d_{1}\leq \sqrt{d},d_{1}|d\}.  \label{A11}
\end{eqnarray}

Applying Theorem 6, we see that Eq. (\ref{A10}) is equivalent to
\begin{equation}
\chi _{AB}\leq d^{\prime }+d/d^{\prime }.   \label{A12}
\end{equation}

Next, we prove that $\chi _{AB}\geq d^{\prime }+d/d^{\prime }$, then Theorem 7 follows. For simplicity of notation, we let $d/d^{\prime }=d^{\prime \prime }.$

\begin{figure}
\includegraphics[width=8cm]{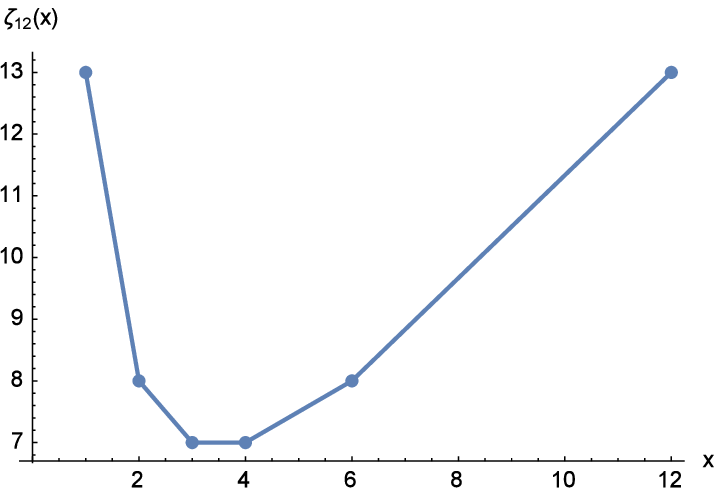}
\caption{Plot of $\varsigma_{12}(x)$ in Eq. (\ref{A15}). All $x\in [3,4]$ reach the minimum $\varsigma_{12}(3)=\varsigma_{12}(4)=7.$}
\includegraphics[width=8cm]{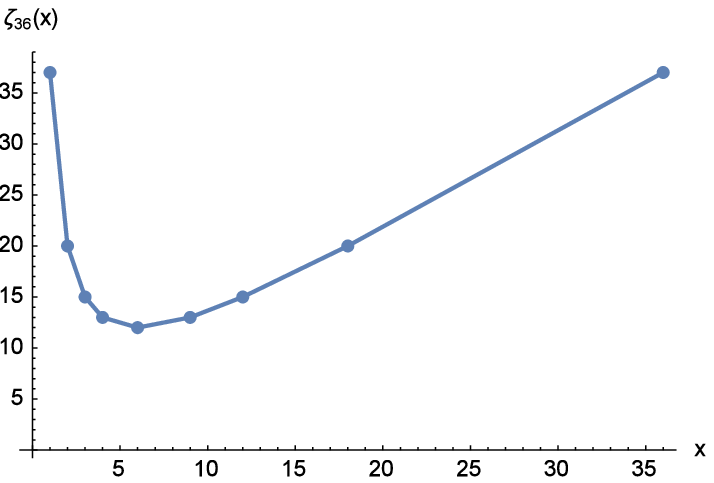}
\caption{Plot of $\varsigma_{36}(x)$ in Eq. (\ref{A15}). Only one value $x=6$ reaches the minimum $\zeta _{d}(6)=12.$}
\end{figure}

\emph{Lemma 1}(\cite{EJC-2006-Meshulam}). Let $d_{1}<d_{2}$ be two
consecutive divisors of $d$. If $d_{1}\leq |$supp$f|\leq d_{2}$ then
\begin{equation}
|\text{supp}\widehat{f}|\geq \frac{d}{d_{1}d_{2}}(d_{1}+d_{2}-|\text{supp}%
f|).  \label{A13}
\end{equation}

Adding $|$supp$f|$ to both sides of Eq. (\ref{A13}), we get
\begin{equation}
|\text{supp}f|+|\text{supp}\widehat{f}|\geq \frac{d}{d_{1}}+\frac{d}{d_{2}}%
+(1-\frac{d}{d_{1}d_{2}})|\text{supp}f|.  \label{A14}
\end{equation}

Define the function
\begin{equation}
\zeta _{d}(x)=\frac{d}{d_{1}(x)}+\frac{d}{d_{2}(x)}+[1-\frac{d}{%
d_{1}(x)d_{2}(x)}]x,  \label{A15}
\end{equation}
where $x\in \lbrack 1,d],$ $d_{1}(x)$ is the greatest divisor of $d$
satisfying $d_{1}(x)\leq x$, $d_{2}(x)$ is the least divisor of $d$
satisfying $d_{2}(x)\geq x.$ If $x=q$ being a positive integer and $q|d,$
then $d_{1}(q)=d_{2}(q)=q.$ $\zeta _{d}(x)$ has the obvious properties below.

(7.1). $\zeta _{d}(q)=\zeta _{d}(\frac{d}{q})=q+\frac{d}{q}$ when $q|d.$

(7.2). $\zeta _{d}(x)=d^{\prime }+d^{\prime \prime },x\in \lbrack d^{\prime
},d^{\prime \prime }].$

(7.3). $\zeta _{d}(x)$ is linear with respect to $x$ when $x\in \lbrack d_{1},d_{2}]$ and  $d_{1}<d_{2}$
are two consecutive divisors of $d.$

(7.4).
\begin{equation}
 \begin{cases}
 1-\frac{d}{d_{1}(x)d_{2}(x)}<0, \text{when} \ x\in (1,d^{\prime }); \\
 1-\frac{d}{d_{1}(x)d_{2}(x)}=0, \text{when} \ x\in (d^{\prime },d^{\prime \prime }); \\
 1-\frac{d}{d_{1}(x)d_{2}(x)}>0, \text{when} \ x\in (d^{\prime },d].
\end{cases}
\end{equation}

Consequently, $\zeta _{d}(x)$ decreases when $x$ increases in $[1,d^{\prime
}],$ $\zeta _{d}(x)$ increases when $x$ increases in $[d^{\prime \prime
},d], $ and $\zeta _{d}(x)$ keeps constant when $x$ increases in $[d^{\prime
},d^{\prime \prime }].$ It follows that $\zeta _{d}(x)\geq d^{\prime
}+d^{\prime \prime }$ and the lower bound $d^{\prime }+d^{\prime \prime }$ is reached only when $x\in [d^{\prime },d^{\prime \prime }]$. Also, $\zeta_{d}(x)$ is a convex function. When $d$ is a square number, then $d^{\prime
}=d^{\prime \prime },$ $d=d^{\prime 2},$ only one value $x=d^{\prime
}$ reaches the minimum $\zeta _{d}(d^{\prime})=2d^{\prime}$. When $d$ is not a square number, then $d^{\prime}<d^{\prime \prime },$ all $x\in[d^{\prime},d^{\prime \prime }]$ reach the minimum $\zeta _{d}(d^{\prime})=\zeta _{d}(d^{\prime \prime})=d^{\prime}+d^{\prime \prime}$. We plot $\zeta _{12}(x)$ (12 is not a square number) and $\zeta _{36}(x)$ (36 is a square number) in Figure 2 and Figure 3.

Returning to Eq. (\ref{A14}), we get that
\begin{equation}
|\text{supp}f|+|\text{supp}\widehat{f}|\geq d^{\prime }+d^{\prime \prime },  \label{A16}
\end{equation}
this certainly implies that
\begin{equation}
\chi _{AB}\geq d^{\prime }+d^{\prime \prime }. \label{A17}
\end{equation}

Combining Eqs. (\ref{A12},\ref{A17}), Theorem 7 then follows.


%

\end{document}